\documentclass[
reprint,
showpacs,
nofootinbib,
amsmath,amssymb,
aps,
pra,
floatfix,
]{revtex4-1}

\usepackage{graphicx}
\usepackage{dcolumn}
\usepackage{bm}
\usepackage[colorlinks=true, citecolor=blue, urlcolor=blue ]{hyperref}
\usepackage{amsmath, amssymb}
\usepackage{bbm}
\usepackage{pxfonts,txfonts}

\newcommand{\bq}{\begin{equation}}
\newcommand{\eq}{\end{equation}}
\newcommand{\bqs}{\begin{equation*}}
\newcommand{\eqs}{\end{equation*}}
\newcommand{\ba}{\begin{array}}
\newcommand{\ea}{\end{array}}
\newcommand{\bas}{\begin{array*}}
\newcommand{\eas}{\end{array*}}
\newcommand{\bqa}{\begin{eqnarray}}
\newcommand{\eqa}{\end{eqnarray}}
\newcommand{\bqas}{\begin{eqnarray*}}
\newcommand{\eqas}{\end{eqnarray*}}

\newcommand{\x}{\bm{x}}
\newcommand{\y}{\bm{y}}
\newcommand{\e}{\bm{e}}
\newcommand{\sig}{\bm{\sigma}}
\newcommand{\D}{\mathcal{D}}

\newcommand{\CQ}{\mathcal{CQ}}

\DeclareMathOperator{\diag}{diag}

\newcommand{\etal}{\textit{et al.}}

\newcommand{\ran}{\rangle}
\newcommand{\lan}{\langle}

\newcommand{\eeqref}{Eq.~\eqref}
\newcommand{\ccite}{Ref.~\cite}

\newtheorem{proposition}{Proposition}

\begin{document}


\title{Maximally discordant separable two qubit $X$ states}
\author{Swapan Rana}
 \email{swapanqic@gmail.com}
\author{Preeti Parashar}
\email{parashar@isical.ac.in}
\affiliation{Physics and Applied Mathematics Unit, Indian Statistical Institute, 203 B T Road, Kolkata, India}
\date{\today}

\begin{abstract} In a recent article S. Gharibian [\href{http://dx.doi.org/10.1103/PhysRevA.86.042106}{Phys. Rev. A {\bf 86}, 042106 (2012)}] has conjectured that no two qubit separable state of rank greater than two could be maximally non classical (defined to be those which have normalized geometric discord $1/4$) and asked for an analytic proof. In this work we prove analytically that among the subclass of $X$ states, there is a unique (up to local unitary equivalence) maximal separable state of rank two. Partial progress has been made towards the general problem and some necessary conditions have been derived.
\end{abstract}


\maketitle


As it is well known, the basic tasks in quantum information theory are mainly characterization, quantification and possible applications of quantum correlations. Of these, the characterization part is naturally the elementary step where different notions of \emph{quantumness} emerge from different perspectives. It is always interesting to characterize the states having maximum quantumness (subsequently depends on both the notion and the measure of quantumness), because in general, different notions induce different ordering on the state space. For example, given two entangled measures $E_1, E_2$, there are states $\rho_1,\rho_2$ such that $E_1(\rho_1)>E_2(\rho_1)$, but $E_1(\rho_2)<E_2(\rho_2)$ \cite{Eisert&Plenio.JMO.1999}. Thus, maximally entangled states with respect to (w.r.t.) $E_1$ need not be the same w.r.t. $E_2$. As a result, historically whenever a new measure was proposed, this question was raised subsequently. For some classic examples, see \cite{Eisert&Plenio.JMO.1999,Verstraete&O.PRA.2001,Miranowicz&Grudka.PRA.2004}.

Naturally, the recently introduced quantum discord \cite{Ollivier&Zurek.PRL.2001}, or its well studied variant, the geometric discord  \cite{Dakic&O.PRL.2010,Luo&Fu.PRA.2010,Rana&Parashar.PRA.2012,Hassan&O.PRA.2012}  should not be any exception. It is known that the \emph{usual} geometric discord reaches its maximum only on \textit{maximally entangled} states ($|\Psi\ran=\sum_i|ii\ran/\sqrt{\dim}$) \cite{Luo&Fu.PRL.2011}. However, a much advertised distinctive feature of quantum discord is that it can be non-zero even for separable states. Therefore an obvious question would be: \textit{what is the maximum discord for separable states}? To our knowledge, the first general bound on entropic discord \cite{Ollivier&Zurek.PRL.2001} for separable states was $\delta_A\le\min\{S_A,S_B,\mathcal{I}(A:B)\}$, given by A. Datta (see pp. 40 of \cite{ADattaPhDThesisAr08}). Since then, a vast literature has appeared for characterization of maximally discordant states --- applying both analytical \cite{Gharibian&O.IJQI.2011,Okrasa&Walczak.EPL.2012,Adhikari&Banerjee.PRA.2012} and numerical techniques \cite{Lang&Caves.PRL.2010,Galve&O.PRA.2011,Girolami&Adesso.PRA.2011,Batle&O.JPA.2011,Batle&O.PRA.2012}, mainly for two qubits. Very recently, Gharibian \cite{Gharibian.PRA.2012} has proved analytically that \emph{among rank two  separable states of two qubits}, the maximum value of normalized geometric discord is $1/4$ and conjectured that no separable two qubit state of higher rank could achieve this value. The aim of the present work is to explore this conjecture.

Before proceeding further, let us define the relevant quantities. The main object, the geometric discord (GD), for an $m\otimes n$ state is defined by (normalized to have maximum value $1$) \begin{equation}
\label{Def: GD} \D(\rho)=\frac{m}{m-1}\min_{\chi\in\Omega_0}\|\rho-\chi\|^2,
\end{equation}
where $\Omega_0$ is the set of \emph{zero-discord} or \emph{classical-quantum} ($\CQ$) states (given by $\sum p_k|\psi_k\ran_A\lan \psi_k|\otimes\rho_k^B$) and $\|X\|^2=$ Tr$(X^\dagger X)$ is the Frobenius or Hilbert-Schmidt norm. 
Consider an arbitrary two qubit state in the Bloch form 
\begin{subequations}
\label{Bloch2qs}
\begin{align}
\rho&=\frac{1}{4}\left[I\otimes I+\x^t\sig\otimes I+I\otimes\y^t\sig+\sum T_{ij}\sigma_i\otimes \sigma_j\right]\\
&:=(\x,\y,T).
\end{align}
\end{subequations} Then its GD can be calculated analytically \cite{Dakic&O.PRL.2010},
\begin{equation}
\label{Formula: GD} \D(\rho)=\frac{1}{2}\left[\|\x\|^2+\|T\|^2-\lambda_{\max}(\x\x^t+TT^t)\right],
\end{equation}
with the optimal $\CQ$ state given by $\chi=(\e^t\x\e,\y,\e\e^tT)$ \cite{Dakic&O.PRL.2010,Miranowicz&O.PRA.2012} and $\e$ being the eigenvector of \begin{equation}
\label{def:G} G:=\x\x^t+TT^t
\end{equation}
corresponding to the maximum eigenvalue of $G$. Denoting by $\lambda_i^{\downarrow}(X)$ and $\lambda_i^{\uparrow}(X)$ the eigenvalues (counted multiplicities) of $X$ in non-increasing and non-decreasing order respectively, the conjecture thus becomes
\begin{equation}
\label{Gharibean.conjecture}\max_{\{\text{ Separable }\rho\}}\sum_{i=2}^3\lambda_{i}^\downarrow(G)=\frac{1}{2}
\end{equation}

Although geometric discord is necessarily a quantum resource, at least in some restricted scenario \cite{Dakic&O.NP.2012,PHorodecki&O.arXive.2013},  we must emphasize that irrespective of usefulness of geometric discord, this problem is interesting in its own right. As evident from \eeqref{Gharibean.conjecture}, the problem can be cast as an optimization problem among separable states, without any relation to discord. Also, we note the close resemblance with important inequalities:
\begin{subequations}
\begin{align}
\|T\|_1&:=\sum_{i=1}^3\sqrt{\lambda_i^\downarrow(TT^t)}\le 1\label{separable.condition.T}\\
M(\rho)&:=\sum_{i=1}^2\lambda_i^\downarrow(TT^t)\le 1.
\end{align}
\end{subequations}
The first inequality is a necessary condition for separability for two qubit states \cite{Vicente.QIC.2007}. The last one is a sufficient condition for satisfaction of CHSH inequality for two qubit states \cite{RMPHorodecki.PLA.1995}. 
 
\textit{Maximally non classical separable two qubit $X$ states.} As it is almost customary, first we will consider the $X$ states. This family includes Bell diagonal states, Werner states and corresponds to many physical systems, e.g., the Ising and XY models etc. A detailed study of discord of $X$ states has  been carried out recently in \ccite{Bellomo&O.PRA.2012}. However, the analysis there is unnecessarily complicated due to consideration of completely irrelevant phases. Also, the present question was out of their purview.
\begin{proposition}
\label{Prop:maxDX} The maximum of $\D$ among two qubit separable $X$-states is $1/4$. Moreover, all such maximal states have rank 2.
\end{proposition}
In computational basis, two qubit $X$-states are given by \begin{equation}
\label{Xform}\rho=\left(\begin{array}{cccc}
a & 0 & 0 &p\\
 0    &b&q&0\\
0    &q&c&0\\
p&0&0&d
\end{array}\right),
\end{equation}
where, without loss of generality, we have taken all entries non-negative. Because, the local unitary (LU) transformation $$|0\ran_k\to
\exp\left(i\frac{-\theta_p+(-1)^k\theta_q}{2}\right)|0\ran_k$$ will drive out the phases of $p$, $q$, and neither $\D$ nor rank changes under LU.
 
The requirement $\rho \ge 0$ gives the constraints $p^2\le ad$ and $q^2\le bc$. We also need the separability constraints, i.e., positivity of partial transposition (PPT). Noting that the partial transposition just interchanges $p$ and $q$, it follows that $\rho$ represents a separable state iff \begin{equation}
\label{condsepX}\max\{p,q\}\le\min\{\sqrt{bc},\sqrt{ad}\}.
\end{equation}

With explicit calculation, we have $\x=(0,0,a+b-c-d)$ and $G=\diag\{4(p+q)^2,4(p-q)^2,2(a-c)^2+2(b-d)^2\}$. Therefore, 
\begin{subequations}\label{derivationDforX}\begin{align}
\sum_{i=1}^2\lambda_i^{\uparrow}(G)&\le 8(p^2+q^2)\label{a}\\
&\le 16\min\{ad,bc\},\label{b}
\end{align}
\end{subequations}
where equality occurs in Eq. \eqref{a} iff \begin{equation}
\label{restp+q} 4(p+q)^2\le 2(a-c)^2+2(b-d)^2
\end{equation}
and equality occurs in Eq. \eqref{b} iff \begin{equation}
\label{restpq} p=q=\min\{\sqrt{ad},\sqrt{bc}\}
\end{equation}
As we are seeking for maximum, it follows from Eq. \eqref{b} that the maximum occurs iff \bq\label{ad=bc} ad=bc,\eq
and the maximum value in Eq. \eqref{derivationDforX} becomes $\max\{16ad\} $ subject to 
\begin{subequations}\label{finalxconstraint}\begin{align}
ad=bc&=\frac{1}{8}\left[(a-c)^2+(b-d)^2\right]\label{aa}\\
&a+b+c+d=1.\label{bb}
\end{align}
\end{subequations}
We show in Appendix that this maximum occurs at $a=b=(2\pm\sqrt{2})/8,c=d=1/(32a)$ and hence maximum possible value of $\D$ is $1/4$. 

We also note that the conditions \eqref{restpq} and \eqref{ad=bc} were necessary to achieve this maximum. Thus, it is necessary that the state
 has rank 2 and up to LU, the unique separable $X$-state having the maximum $\D$ is given by \begin{equation}\label{finalMaxX}
\rho=\frac{1}{4\sqrt{2}}\begin{pmatrix}
\sqrt{2}+1&0&0&1\\
0&\sqrt{2}+1&1&0\\
0&1&\sqrt{2}-1&0\\
1&0&0&\sqrt{2}-1
\end{pmatrix}
 \end{equation}  \hfill $\blacksquare$

Quite surprisingly, the authors of Ref. \cite{Girolami&Adesso.PRA.2011} have obtained this state numerically as the optimal one, starting from a rank two $X$ state. On the other hand, the author of Ref. \cite{Gharibian.PRA.2012} has given the unique (up to LU) optimal state among rank two separable state as \begin{equation}
\label{GharibianMaxState} \sigma= \frac{1}{2}\left(|00\ran\lan 00|+|+1\ran\lan +1|\right)=\frac{1}{4}\begin{pmatrix}
2&0&0&0\\0&1&0&1\\0&0&0&0\\0&1&0&1
\end{pmatrix}.
\end{equation}

Although, $\sigma$ apparently does not looks like an $X$ state, we note that $\rho$ and $\sigma$ are LU equivalent, namely $\sigma=(U\otimes V)\rho(U\otimes V)^\dagger$ with $$U=\left(
\begin{array}{cc}
 \frac{1+\sqrt{2}}{\sqrt{4+2\sqrt{2}}} & \frac{1-\sqrt{2}}{\sqrt{4-2\sqrt{2}}} \\
 \frac{1}{\sqrt{4+2\sqrt{2}}} & \frac{1}{\sqrt{4-2\sqrt{2}}}
\end{array}
\right),\quad V=\frac{1}{\sqrt{2}}\left(
\begin{array}{cr}
 1 & 1 \\
 1 & -1
\end{array}
\right).$$

Now let us give\\ \emph{Some necessary conditions for maximally discordant separable states}.
\begin{proposition}
\label{Prop:xcantbe0} No two qubit separable state with $\x=\mathbf{0}$, or $TT^t=\lambda^2I$ could be maximally discordant.
\end{proposition}
By Proposition~\ref{Prop:maxDX}, a maximally discordant separable state must have $\D(\rho)\ge 1/4$. 

Now, from \eeqref{separable.condition.T}, a necessary condition for separability is $\sum\sigma_i(T)\le 1$ \cite{Vicente.QIC.2007}. So, assuming the singular values of $T$ as $a,b,c \ge 0$, we must have \begin{subequations}
\begin{align}
a+b+c&\le 1\\ a^2+b^2+c^2-\max\{a^2,b^2,c^2\}&\ge \frac{1}{2}
\end{align}
\end{subequations} 
which is clearly impossible, as the maximum of $a^2+b^2+c^2-\max\{a^2,b^2,c^2\}$ subject to the constraints $a+b+c\le 1$ and non-negative $a,b,c$ is $2/9<1/2$. 

The second assertion follows by noticing that the eigenvalues of $G$ then become $\{\|x\|^2+\lambda^2,\lambda^2,\lambda^2\}$.  \hfill $\blacksquare$

Remark: The \emph{separability} condition can not be ignored in proposition \ref{Prop:xcantbe0}, as for the Werner state
\[ \rho_w=p|\Psi^-\ran\lan\Psi^-|+\frac{(1-p)}{4}I\] where $|\Psi^-\ran=(|01\ran-|10\ran)/\sqrt{2}$, we have $\x=\mathbf{0}$ and $\D=p^2$ thereby $\D>1/2$ whenever $p>1/\sqrt{2}$. The conjecture in \eeqref{Gharibean.conjecture} predicts the stronger result that $\D(\rho)>1/2$ implies the state is surely entangled, irrespective of $\x$. Note that this result also prohibits the separable Bell-diagonal states to be maximally discordant \cite{Girolami&Adesso.PRA.2011}.

The separability condition in the conjecture is really important, even for existence of extrema.
\begin{proposition}
\label{Prop:Dhasnotmaximum} The function $\D$ has no maximum among rank 2 two qubit states.
\end{proposition}

It is well known that the maximum of $D(\rho)$ is 1 and  attained only at pure maximally entangled states (i.e., rank one states). So it suffices to show that there is always a rank two state $\rho_\epsilon$ having $D(\rho_\epsilon)=1-\epsilon$. Out of many possibilities, one such rank two state is given by   
\begin{equation}\label{rank2rhoepsilon}
\rho_\epsilon=\left( \frac{1}{2}+\sqrt{\frac{1}{4}-\frac{\epsilon}{3}}\right)|\Psi\ran\lan\Psi|+\left(\frac{1}{2}-\sqrt{\frac{1}{4}-\frac{\epsilon}{3}}\right)|00\ran\lan 00|
\end{equation}
It is easy to verify that the optimal measurement operators $\Pi_{1,2}=(I\pm\sigma_x)/2$ give the required value of $D$. \hfill $\blacksquare$

We note that the state in \eeqref{rank2rhoepsilon} remains entangled for the entire range of $\epsilon\in[0,3/4]$. Also, changing $\epsilon\to 3/4-\epsilon$, it follows that there is always a rank two (entangled) state having $D=1/4+\epsilon$. 

Before finishing, let us mention some of our failed attempts towards this conjecture. We have been able to  prove the conjecture (including the uniqueness), under anyone of the following interrelated assumptions: \begin{itemize}
\item[i.] A maximally discordant separable state (MDSS) $\rho$ must have at least one closest $\CQ$ state as $\rho^A\otimes\rho^B$
\item[ii.] An MDSS should have $\y=\mathbf{0}$
\item[iii.] $G$ has degenerate spectrum for any MDSS
\item[iv.] $G$ is singular for any MDSS
\end{itemize} 
But we do not know why (or how to establish) anyone of these is necessary for MDSS. A quite unpleasant situation occurs while trying to directly solve the optimization problem:

\[\max f(\x,T):=\|\x\|^2+\|T\|^2-\e^tTT^t\e\]
Vanishing of the gradient gives \begin{subequations}
\begin{align}
\frac{\partial f}{\partial \x}=\mathbf{0}&\Rightarrow \x=\e^t\x\e\\
\frac{\partial f}{\partial T}=\mathbf{0}&\Rightarrow T=\e\e^tT
\end{align}
\end{subequations}
These two equations are enough to determine a unique MDSS. But unfortunately, the Hessian matrix is Block-diagonal with block $2(I-\e\e^t)\ge 0$. Thereby we can not guarantee that this is indeed the maxima. 
\appendix*
\section{Proof of the optimization in Proposition \ref{Prop:maxDX}}
We first note that the constraint \eqref{aa} implies $abcd\ne 0$, $a\ne c$, $b\ne d$. Now let us try to parametrize $(a,b,c,d)$ using the constraints.
To absorb the first constraint, without loss of generality, we can take $a=bk,c=dk,k>0$. Then the constraint \eqref{bb} becomes \begin{equation}
\label{b+d=1/k+1}b+d=\frac{1}{k+1},\end{equation}
and we are left with only the following constraint \begin{eqnarray}
 (a-c)^2+(b-d)^2&=&8bc\nonumber\\
\Rightarrow (b-d)^2(k^2+1)&=&8bdk\label{b-d}.\end{eqnarray} 
From \eqref{b+d=1/k+1} and \eqref{b-d}, we get \begin{equation}
\label{bd} 4bd=\frac{k^2+1}{(k+1)^4}
\end{equation}
Noting that $ad=bdk$, we have to find the maximum of the function $$f(k)=\frac{k(k^2+1)}{(k+1)^4}$$ subject to $k>0$. The derivatives are very easy to calculate. Indeed, $f'(k)=0$ only at $k=1$ and $f'(1)=f''(1)=f'''(1)=0$, but $f''''(1)=-3/16<0$. Hence, the unique maximum occurs at $k=1$. From \eqref{b+d=1/k+1} and \eqref{bd}, this corresponds to the solution $a=b=(2\pm\sqrt{2})/8,c=d=1/(32a)$. \hfill $\blacksquare$

\end{document}